\documentclass{ecc}
\usepackage{amsmath}
\usepackage{comment}
\usepackage{graphicx}
\usepackage{amsfonts}
\usepackage{enumitem}  
\usepackage{booktabs}
\usepackage{multirow}
\usepackage{tabularx}
\usepackage{array}
\usepackage{makecell}
\usepackage{xcolor}  
\usepackage{url}
\title{Hierarchical Reinforcement Learning with Low-Level MPC for Multi-Agent Control}

\author{Max Studt$^1$, Georg Schildbach$^1$}

\begin{document}
\maketitle

{
  \renewcommand{\thefootnote}{}%
  \footnotetext[1]{$^1$Institute for Electrical Engineering in Medicine of the University of Luebeck, Germany, (\emph{\{m.studt, georg.schildbach\}@uni-luebeck.de})}
}

\begin{abstract}
Achieving safe and coordinated behavior in dynamic, constraint-rich environments remains a major challenge for learning-based control. Pure end-to-end learning often suffers from poor sample efficiency and limited reliability, while model-based methods depend on predefined references and struggle to generalize. We propose a hierarchical framework that combines tactical decision-making via reinforcement learning (RL) with low-level execution through Model Predictive Control (MPC). For the case of multi-agent systems this means that high-level policies select abstract targets from structured regions of interest (ROIs), while MPC ensures dynamically feasible and safe motion. Tested on a predator–prey benchmark, our approach outperforms end-to-end and shielding-based RL baselines in terms of reward, safety, and consistency, underscoring the benefits of combining structured learning with model-based control.$^2$
\end{abstract}

\footnotetext[2]{Demonstration video available at:\\ \url{https://www.youtube.com/watch?v=n9BBm2uZHlk}} 

\section{Introduction}
Reinforcement learning (RL) has emerged as a powerful tool for solving complex decision-making problems, offering the ability to learn control policies from interaction with an environment. However, deploying RL in safety-critical applications remains a significant challenge. Standard end-to-end learning approaches often lack formal safety guarantees and suffer from poor sample efficiency, especially when required to satisfy hard physical constraints such as collision avoidance, actuator saturation, or operational safety limits. Learning both task behavior and constraint satisfaction simultaneously can lead to unstable or slow convergence and, in real-world systems, may even render training infeasible~\cite{zhao2023statewisesafereinforcementlearning, wachi2025provableapproachendtoendsafe}.

At the same time, purely model-based control methods, such as MPC, while effective in enforcing constraints and guaranteeing safe execution, typically rely on externally defined reference trajectories. In static or well-structured settings, such references can be manually designed. However, in complex, dynamic, or partially known environments, designing meaningful reference trajectories a priori is challenging. As a result, model-based approaches struggle to adapt to changing goals, task objectives, or environment layouts without external guidance. This limits their autonomy and applicability in many real-world scenarios.

These complementary limitations make a hybrid approach particularly attractive: RL can provide high-level, adaptive intent in the form of target selection or task decisions, while model-based control can ensure safety, constraint satisfaction, and physical feasibility during execution. Consider, for example, a fleet of autonomous delivery drones operating in a changing urban environment. While MPC can ensure collision-free trajectories and respect dynamic constraints such as battery limits or no-fly zones, it cannot decide where to go or how to prioritize deliveries without predefined references. Conversely, an RL-based policy can flexibly decide on routing and target selection based on learned objectives, but may fail to produce safe low-level motions if trained in an end-to-end fashion. The integration of both paradigms in a hierarchical structure allows strategic reasoning and adaptive behavior on top of a reliable, constraint-respecting execution layer, offering a promising path toward safe and generalizable learning-based control.

An especially challenging application domain for RL arises in multi-agent systems, where agents must coordinate under coupled dynamics and safety constraints, often in complex environments. Multi-agent reinforcement learning (MARL) in such settings exacerbates the sample inefficiency and safety limitations of single-agent RL due to the increased dimensionality, non-stationarity, and the need for implicit coordination \cite{NING202473}. Moreover, safety violations by individual agents can propagate through the system, further increasing the difficulty of ensuring constraint satisfaction during both training and execution.

This paper addresses precisely these challenges. We propose a hierarchical MARL framework that decouples high-level decision-making from low-level control, combining the flexibility of learned coordination strategies with the safety and reliability of model-based execution. Through structured target selection and decentralized constraint handling, our method aims to enable safe, efficient, and robust behavior in dynamic, multi-agent systems.

\section{Related Work and Contribution}
The idea of combining RL with model-based control has been explored from multiple angles, particularly to mitigate safety risks and to improve the sample efficiency in complex decision-making tasks. Prior work can broadly be divided into single-agent and multi-agent approaches, each with different assumptions and integration strategies.

In the single-agent setting, several methods aim to tune robust MPC schemes within a constrained RL framework, for example by learning cost weights and constraint tightenings from data, while keeping the overall structure hand-crafted and modular \cite{9198135, Robust2}. 
Other approaches incorporate uncertainty estimates through Gaussian Process models and leverage MPC for safe exploration under probabilistic guarantees \cite{koller}. 
Formal safety guarantees during policy updates have also been obtained using Lyapunov-based criteria \cite{berkenkamp2017safemodelbasedreinforcementlearning}. 
Additional methods introduce safety filters or shielding mechanisms that override unsafe actions at execution time \cite{ZeilingerSafteyFilter, li2020robustmodelpredictiveshielding}. 
A broader taxonomy and classification of these approaches is available in a recent survey work \cite{reiter2025synthesismodelpredictivecontrol}.

Extending such ideas to multi-agent systems presents further challenges due to inter-agent coupling and the curse of dimensionality. 
One line of work uses learned policies to warm-start MPC optimizers during deployment, thereby enhancing safety without changing the learned decision structure \cite{deepSafe}. 
Other methods employ distributed MPC as a safety shield that filters actions in cooperative navigation tasks and penalizes unsafe deviations \cite{10964685}. 
Hierarchical architectures have also been proposed, in which a robust multi-agent policy handles high-level coordination under observation uncertainty, while a low-level MPC controller ensures safe execution through control barrier functions \cite{zhang2024safetyguaranteedrobustmultiagent}.

\subsubsection*{Our Contribution}
\begin{figure}[ht]
  \centering
  \includegraphics[width=0.335\textwidth]{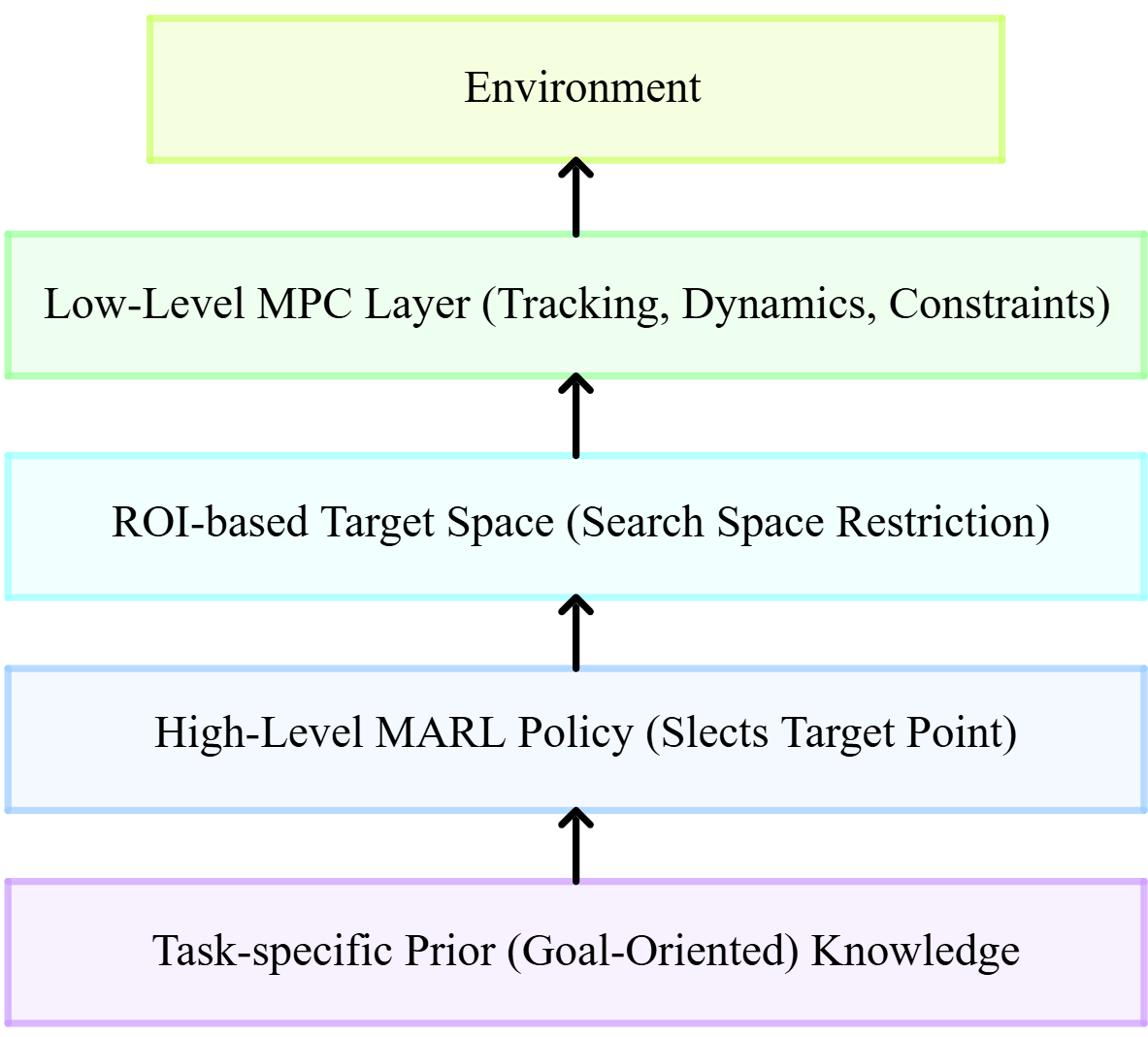}
  \caption{Overview of our hierarchical decision-making and control framework. The high-level MARL policy selects discrete targets from a structured ROI, informed by task-specific prior knowledge. A low-level MPC controller then tracks the selected target under dynamics and safety constraints.}
  \label{fig:control-architecture}
\end{figure}

In contrast to previous approaches, we propose a new RL framework (shown in Fig. \ref{fig:control-architecture}) that does not treat MPC as a passive safety layer. Instead, the two are combined in a hierarchical order, during learning and execution, attempting to decouple the strategic decision making via RL from the low-level system control via MPC. The key contributions are as follows:

(\textit{i}) A hierarchical RL-MPC control framework that uses task-specific prior knowledge to structure each agent’s action space into goal-relevant regions of interest (ROIs) around likely coordination targets. Here, the policy outputs continuous target points from these ROIs, which a decentralized MPC then tracks under constraints. This improves sample efficiency and stability, especially in sparse-reward settings.

(\textit{ii}) Application of this concept to cooperative MARL, demonstrated on a predator-prey benchmark with discrete-continuous hybrid actions, nonlinear polar coordinate transformations and minimal reward shaping.

(\textit{iii}) Experimental evaluation against end-to-end learning and shielding-based MPC baselines, demonstrating superior performance in reward, safety, and behavioral consistency.

\section{Background and Preliminaries}

In this section, we lay out the formal foundations for a hierarchical control architecture that couples learned decision-making with model-based execution. We start by reviewing cooperative MARL under the centralized-training, decentralized-execution (CTDE) paradigm, which can be cast as a decentralized partially observable Markov decision process (Dec-POMDP) \cite{DEC-POMDP}. We then introduce a generic continuous-state, continuous-action agent model and formulate the decentralized MPC problem used for low-level target tracking. 

\subsection{Cooperative Multi-Agent Reinforcement Learning}
We model cooperative decision making as a decentralized partially observable Markov decision process (Dec-POMDP) for \(N\) agents \cite{DEC-POMDP}. Formally, a Dec-POMDP can be expressed as a tupel:
\[
M = \bigl(\,\mathcal S,\,\mathcal A,\,\mathcal O,\,P,\,\Omega,\,R,\,n,\,\gamma\bigr),
\]
where \(n\) is the number of agents and \(\mathcal S\) the global state space. Defining \(\mathcal{I}=\{1,\ldots,n\}\) as the agent set, \(\mathcal A_i\) represents the action space of agent \(i\) (with joint action space \(\mathcal A = \prod_{i\in\mathcal I}\mathcal A_i\)), and \(\mathcal O_i\) the observation space of agent \(i\) (with joint observation space \(\mathcal O = \prod_{i\in\mathcal I}\mathcal O_i\)). The state transition function
\[
P(s' \mid s, \mathbf a): \mathcal S \times \mathcal A \times \mathcal S \rightarrow [0,1]
\]
governs the probability of moving from state \(s\in\mathcal S\) to \(s'\in\mathcal S\) under joint action \(\mathbf a\in\mathcal A\), and
\[
\Omega(\mathbf o \mid s', \mathbf a): \mathcal S \times \mathcal A \times \mathcal O \rightarrow [0,1]
\]
defines the probability of observing \(\mathbf o\in\mathcal O\) given \(s'\) and \(\mathbf a\). The shared reward function \(R(s,\mathbf a)\) assigns a numerical reward to each state–action pair, and \(\gamma\in[0,1)\) is the discount factor.

At each time step \(t\), the environment is in state \(s_t \in \mathcal{S}\). Each agent \(i\) receives a private observation \(o_t^i\in\mathcal O_i\) sampled according to the marginal of \(\Omega(\cdot\mid s_t,\mathbf a_{t-1})\), and selects an action \(a_t^i\in\mathcal A_i\) according to its stochastic policy
\[
\pi_i(a_t^i \mid o_t^i; \theta).
\]
The joint action \(\mathbf a_t = (a_t^1, \dots, a_t^N)\) induces a transition \(s_{t+1} \sim P(\cdot \mid s_t, \mathbf a_t)\) and yields the shared reward \(r_t = R(s_t, \mathbf a_t)\). The objective is to learn policy parameters \(\theta\) that maximize the expected discounted return
\[
J(\theta) = \mathbb E\Bigl[\sum_{t=0}^{T-1}\gamma^t\,R(s_t, \mathbf a_t)\Bigr].
\]

\subsection{Decentralized Model Predictive Control}

MPC is an optimal control technique in which, at each time step, an agent solves a finite‐horizon optimal control problem, applies the first control action, and then shifts the horizon forward in a receding‐horizon fashion.  In a decentralized multi‐agent setting, each agent \(i\in\mathcal I\) independently computes an optimal control sequence 
\(\mathbf u_i^*=[u_{i,0},\dots,u_{i,N-1}]\)
over a prediction horizon \(N\), subject to its own discrete‐time dynamics 
\(\;x_{i,k+1}=f(x_{i,k},u_{i,k}),\;x_{i,0}=x_i(t),\)
and to constraints that ensure safety and feasibility in the presence of other agents and obstacles.

A general finite‐horizon MPC problem which will be used later for agent \(i\) is then
\begin{equation}\label{eq:mpc-problem}
\begin{aligned}
\min_{\mathbf u_i,\mathbf x_i,\mathbf \xi_i} \
  &\sum_{k=0}^{N-1}\ell(x_{i,k},u_{i,k})
   +\sum_{j=1}^{m_s}\sum_{k=0}^{N}w_j\,\xi_{i,j,k}^q,\\
\text{s.t.} \ \ 
  &x_{i,k+1}=f(x_{i,k},u_{i,k}), \ k=0,\dots,N-1,\\
  &h_l(x_{i,k},u_{i,k})\le0, \ l=1,\dots,m_c,\;k=0,\dots,N-1,\\
  &g_j(x_{i,k})+\xi_{i,j,k}\ge0, \ j=1,\dots,m_s,\;k=0,\dots,N,\\
  &\xi_{i,j,k}\ge0.  
\end{aligned}
\end{equation}

Here:
\begin{itemize}
\item  \(f\colon\mathcal X_i\times\mathcal U_i\to\mathcal X_i\) defines the agent’s dynamics,  
\item  \(\ell\colon\mathcal X_i\times\mathcal U_i\to\mathbb R_{\ge0}\) is the stage cost,  
\item  \(h_l\colon\mathcal X_i\times\mathcal U_i\to\mathbb R\) encodes the \(m_c\) \emph{hard constraints} that must be satisfied at all times (e.g., actuator limits, strict collision avoidance).  
\item  \(g_j\colon\mathcal X_i\to\mathbb R\) defines the \(m_s\) \emph{soft constraints}, such as minimum distances to other agents or obstacles. These requirements are relaxed through slack variables to preserve feasibility.  
\item  \(\xi_{i,j,k}\ge0\) are slack variables penalized by weights \(w_j>0\) and exponent \(q\ge1\).  
\end{itemize}

After solving \eqref{eq:mpc-problem}, agent $i$ applies $u_{i,0}$, observes the new state, and repeats the optimization at the next time step. Through safety constraints that are evaluated relative to neighbors, the planned motion of other agents affects $i$’s feasible set. This fully decentralized receding-horizon scheme lets each agent react locally while aiming to ensure safety and feasibility under the specified constraints.

\section{Hierarchical RL with Low-Level MPC}
For a broad class of target‐oriented MARL environments, such as autonomous UAV navigation to waypoints \cite{UAV}, multi‐robot area coverage \cite{Coverage}, or cooperative object transport \cite{Transportation}, each agent continuously strives to reach one of a finite set of \(C\) reference positions \(p_{c,\mathrm{target}}\), with \(c \in \{1,\dots,C\}\), in order to fulfill a shared objective. In our hybrid framework, the RL policy is solely responsible for selecting a discrete goal index from the reference set and generating a continuous target-point command within a ROI around that goal. Instead of controlling the agent directly, the RL provides the reference input for an MPC execution layer. The MPC realizes the target point of the RL, by computing a collision-free and dynamically feasible trajectory.

At each time step \(t\), agent \(i\) proceeds as follows:

\begin{enumerate}
    \item The MARL actor chooses a \emph{reference index} \(c\in\{1,\dots,C\}\) and outputs a continuous \emph{target point}
  \[
    \tau_t^i \;\in\; \mathcal T
    = \bigl\{\tau\in\mathbb R^d : \|\tau - p_{c,\mathrm{target}}\|\le r_{\mathrm{ROI}}\bigr\}.
  \]
  Here \(d\) indicates the spatial dimension. This construction guarantees that \(\tau_t^i\) lies within a predefined radius \(r_{\mathrm{ROI}}\) around the \(c\)-th goal \(p_{c,\mathrm{target}}\), effectively reducing the policy’s exploration space and improving training efficiency. Both the region radius \(r_{\mathrm{ROI}}\) and the reference points \(p_{c,\mathrm{target}}\) are specified externally: they serve as tuning parameters to focus the search without unduly constraining the learned tactical behavior.

 \item Agent \(i\) solves, \emph{decentrally}, the general finite‐horizon MPC problem as described in equation~\eqref{eq:mpc-problem}, using the following objective function:
    \[
    \begin{aligned}
      \min_{\{u_{i,k},\,\xi_{i,j,k}\}} \ & 
        \sum_{k=0}^{N-1}\Bigl[(p_{i,k}-\tau_t^i)^\top Q\,(p_{i,k}-\tau_t^i)
        + u_{i,k}^\top R\,u_{i,k}\Bigr] \\
        &\;+\;\sum_{j=1}^{m_s}\sum_{k=0}^{N} w_j\,\xi_{i,j,k}^q.
    \end{aligned}
    \]
    The quadratic term \((p_{i,k}-\tau_t^i)^\top Q\,(p_{i,k}-\tau_t^i)\) drives the agents position \(p_{i,k}\) toward \(\tau_t^i\). The control effort is penalized by the matrix $R$ and the slack variables \(\xi_{i,j,k}\) are used to penalize the violation of minimal safety distances to other agents and obstacles. 
  \item The agent applies only the first control action \(u_{i,0}\), updates its state, and returns to step 1).
\end{enumerate}
By enforcing the ROI at the policy level and restricting the MPC to standard dynamic and safety constraints, we strictly separate \emph{strategic intent} (MARL) from \emph{feasible execution} (MPC). The policy focuses solely on selecting semantically meaningful targets, while MPC generates dynamically consistent control commands without manual trajectory design.

\begin{figure}[ht]
  \centering
  \includegraphics[width=0.24\textwidth]{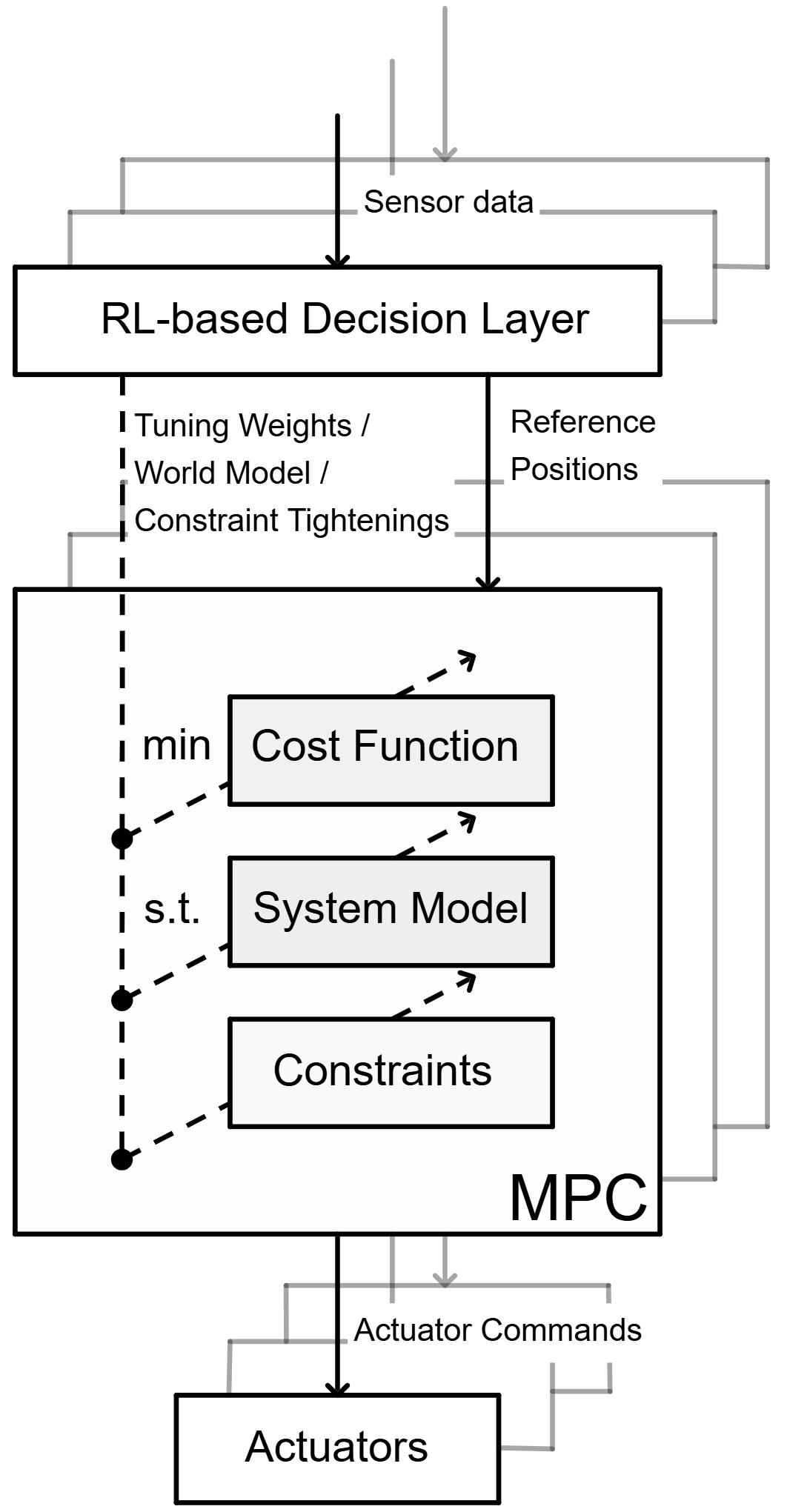}
  \caption{Overview of the RL–MPC hierarchical architecture. The RL policy outputs reference positions for the MPC to track. The framework also permits learning other MPC elements (e.g., cost parameters, constraint margins, or models), enabling adaptive control across multiple layers.}
  \label{fig:Framework}
\end{figure}

\section{Predator-Prey Benchmark}
To evaluate our hybrid RL-MPC approach, we have designed a circular, bounded, two-dimensional continuous‐control predator–prey environment \(\mathcal{X}\) in which two predator agents learn tactical hunting behaviors, and three prey agents follow a simple heuristic evasion policy. 

All predators as well as preys obey discrete‐time double‐integrator dynamics with a hard acceleration bound and state constraints:
\[
x_{k+1} = A\,x_k + B\,u_k,
\quad
\|u_k\|\le u_{\max},
\quad
x_k \in \mathcal X,
\]
where the state and input are defined as
\[
x_k = \begin{bmatrix}x \\ y \\ v_x \\ v_y\end{bmatrix}, 
\quad
u_k = \begin{bmatrix}a_x \\ a_y\end{bmatrix},
\]
with \((x,y)\) and \((v_x,v_y)\) denoting planar position and velocity, and \((a_x,a_y)\) the acceleration commands in \(x\)- and \(y\)-direction.  The matrices \(A\) and \(B\) are given by
\[
A = 
\begin{bmatrix}
1 & 0 & \Delta t & 0\\
0 & 1 & 0 & \Delta t\\
0 & 0 & 1 & 0\\
0 & 0 & 0 & 1
\end{bmatrix}, 
\quad
B =
\begin{bmatrix}
\tfrac12\,\Delta t^2 & 0\\
0 & \tfrac12\,\Delta t^2\\
\Delta t & 0\\
0 & \Delta t
\end{bmatrix}.
\]
To ensure that the preys cannot be caught easily by a single predator, prey agents are endowed with higher motion capabilities than the predators.  Concretely, prey agents have a larger acceleration bound \(u_{\max}^{\rm prey}>u_{\max}^{\rm pred}\) and a higher top speed \(v_{\max}^{\rm prey}>v_{\max}^{\rm pred}\).  This gap enforces that successful capture requires cooperative strategies among the predators. The circular environment eliminates corners in order to prevent trivial tactical encirclement of the preys.

\subsubsection*{ROI-Guided MPC-MARL Approach}

In our predator–prey environment, each predator \(i\) executes the following loop at time \(t\):
\begin{itemize}
    \item[1.] \textit{Tactical goal selection:} The MARL policy picks a prey index  
\[
c \in \{1,\dots,C\}
\]
and outputs two normalized scalars  
\(\;r'\), \(\theta'\in [0,1]\).  
These are then mapped to the actual ROI parameters via  
\[
r = r' \,r_{\mathrm{ROI}},
\quad
\theta = 2\pi\,\theta',
\quad
\tau^i_t = p^{\rm prey}_{c} + r\,[\cos\theta,\;\sin\theta]^\top.
\]
This normalization ensures that the policy always produces outputs in a fixed range \([0,1]\), which we then scale to the desired radius and full angular range. A visualization of this procedure is provided in Fig. \ref{fig:GuidedMPCVisualisation}.
\begin{figure}[ht]
  \centering
  \includegraphics[width=0.34\textwidth]{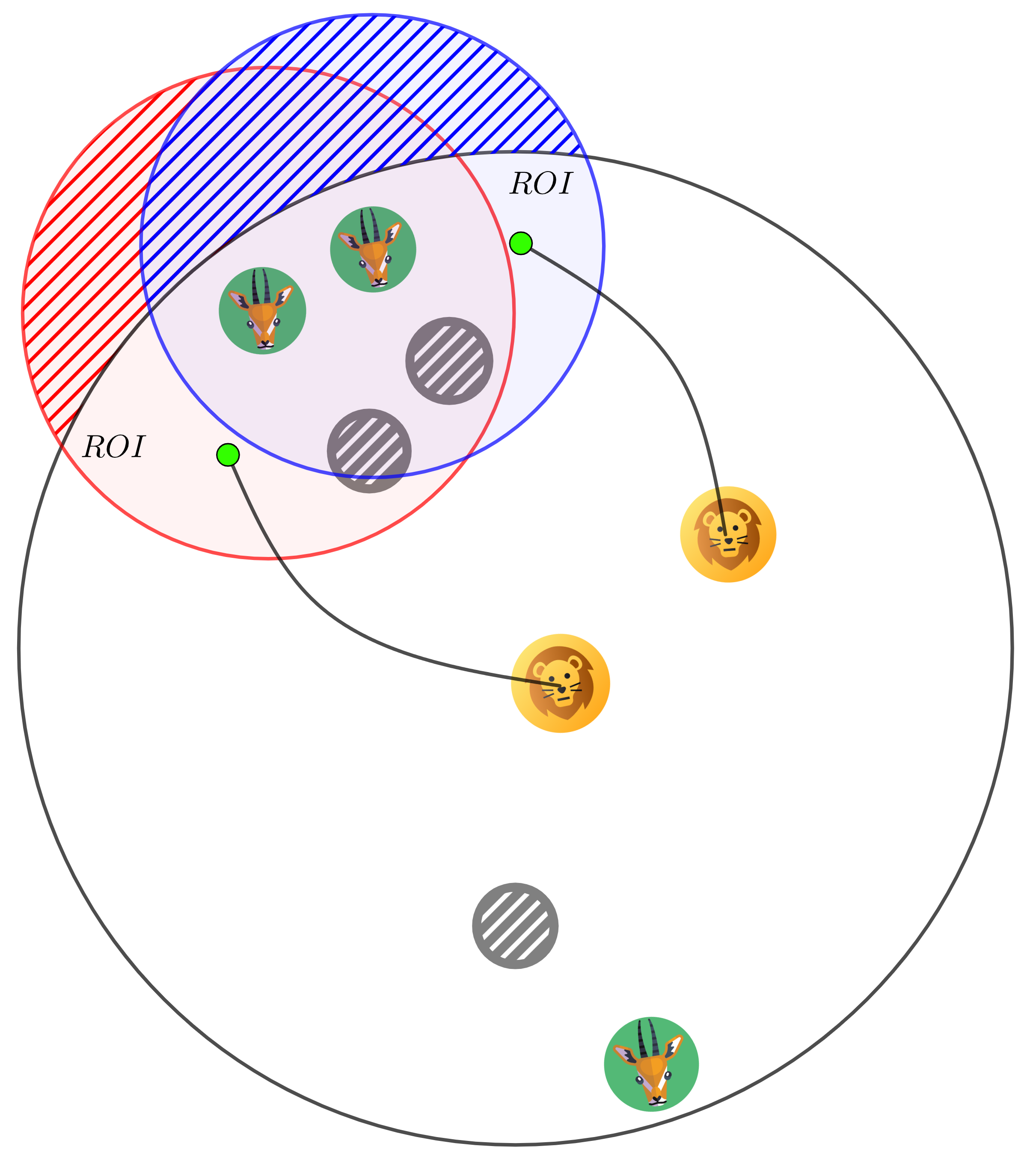}
  \caption{Visualization of our ROI-guided learning approach with target‐point selection within a region of interest in the predator–prey environment. Predators are shown as lion icons, preys as antelope icons, obstacles as grey objects and the target points are shown in green.}
  \label{fig:GuidedMPCVisualisation}
\end{figure}

\item[2.] \textit{Decentralized MPC tracking:}  
Given \(\tau^i_t\), agent \(i\) solves an MPC problem with horizon \(N\): 
\end{itemize}
\begin{equation} \label{eq:mpc-problem-predprey}
\begin{aligned}
  \min_{\{u_{i,k},\xi_{i,k}^{\rm sep},\xi_{i,k,m}^{\rm obs}\}}\, &
    \sum_{k=0}^{N-1}\Bigl[\|p_{i,k}-\tau^i_t\|_{Q}^2 + \|u_{i,k}\|_{R}^2\Bigr]\\
  + &W_{\rm sep}\sum_{k=0}^{N}\Bigl(\tfrac{\xi_{i,k}^{\rm sep}}{\varepsilon_{\rm sep}}\Bigr)^4
    + W_{\rm obs}\sum_{k=0}^{N}\sum_{w=1}^{W}\Bigl(\tfrac{\xi_{i,k,w}^{\rm obs}}{\varepsilon_{\rm obs}}\Bigr)^4,\\
\text{s.t.}\quad
  &x_{i,0}=x_i(t),\quad x_{i,k+1}=A\,x_{i,k}+B\,u_{i,k},\\
  &\|p_{i,k}\|\le r_{\rm world},\quad \\
  & \|p_{i,k}-p_{\rm other}\|^2\ge d_{\rm safe}^2 - \xi_{i,k}^{\rm sep},\\
  &\|p_{i,k}-p^{\rm obs}_w\|^2\ge r_{\rm obs}^2 - \xi_{i,k,w}^{\rm obs},\\
  &\|u_{i,k}\| \le u_{\max},\; \|v_{i,k}\| \le v_{\max},\\
  &\xi_{i,k}^{\rm sep}\ge0, \ \forall k=0,\dots,N, \\
  &\xi_{i,k,w}^{\rm obs}\ge0, \ \forall k=0,\dots,N,\;w=1,\dots,W.
\end{aligned}
\end{equation}
Here:
\begin{itemize}[noitemsep,leftmargin=*]
  \item \(Q,R\) are the standard MPC tracking and control‐effort weight matrices.  
  \item \(\xi^{\rm sep}\), \(\xi^{\rm obs}\) are nonnegative slack variables for separation and obstacle constraints,  
  penalized by weights \(W_{\rm sep},W_{\rm obs}\) and normalized by \(\varepsilon_{\rm sep},\varepsilon_{\rm obs}\).  
  \item \(A,B\) encode the discrete‐time double‐integrator dynamics,  
  and \(r_{\rm world},d_{\rm safe},r_{\rm obs},u_{\max},v_{\max}\) are environment, safety and input/velocity bounds.  
\end{itemize}
\begin{itemize}
\item[3.] \textit{Execution:}  
Only the first optimal control input \(u_{i,0}\) is applied; the state is updated and the loop repeats.
\end{itemize}

\begin{figure}[h]
  \centering
  \includegraphics[width=0.47\textwidth]{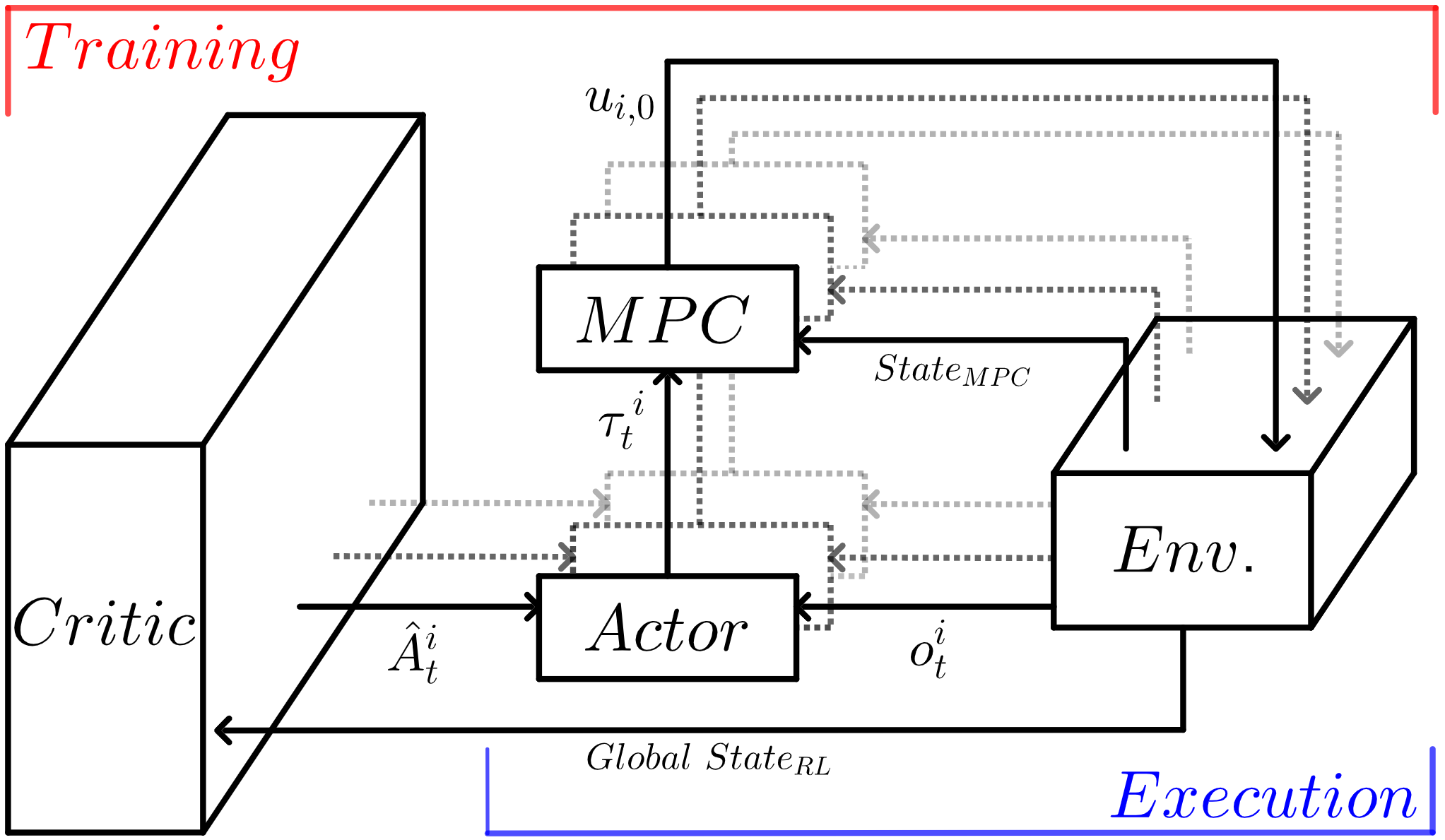}
  \caption{Overview of our hierarchical CTDE architecture. During training (red), a centralized critic computes an advantage estimate \( \hat{A}_t^i \) for each agent \(i\) based on the global state and the agents’ actions. This estimate is used to update the decentralized actor policy. During execution (blue), each actor receives only its local observation \( o_t^i \) and produces tactic parameters, which are passed to a low-level MPC controller that generates continuous control actions \( u_{i,0} \). The environment returns observations and reward signals based on all agents’ actions and states.}
  \label{fig:ctde-architecture}
\end{figure}

Training is conducted using multi-agent proximal policy optimization (MAPPO) within the centralized training with decentralized execution (CTDE) paradigm, as illustrated in Fig.~\ref{fig:ctde-architecture}.

\section{Results}
To quantify the benefits and evaluate the ROI-guided MPC-MARL approach, referred to as \emph{ROI-guided learning} in the following, we compare it against two baseline RL schemes that have shown promise in continuous-control multi-agent tasks:
\begin{itemize}
  \item \textbf{End‐to‐End:} A pure MARL policy directly outputs continuous accelerations \(u_{i,RL}\in[-u_{\max},u_{\max}]^2\).  
  \item \textbf{Shielding MPC:} The same RL policy is used, but its acceleration outputs are post‐processed by a shielding-based MPC, which solves an optimization problem at each time step to enforce collision‐avoidance. The objective of the Shielding MPC formulation is as follows:
  \begin{equation} \label{eq:shielding-mpc-problem-predprey}
        \begin{aligned}
          \min_{\{u_{i,k},\xi_{i,k}^{\rm sep},\xi_{i,k,m}^{\rm obs}\}}\, &
            \|u_{i,RL}-u_{i,0}\|^2_{R_0}\sum_{k=1}^{N-1}\|u_{i,k}\|_{R_S}^2\\
          + W_{\rm sep}&\sum_{k=0}^{N}\Bigl(\tfrac{\xi_{i,k}^{\rm sep}}{\varepsilon_{\rm sep}}\Bigr)^4
            + W_{\rm obs}\sum_{k=0}^{N}\sum_{w=1}^{W}\Bigl(\tfrac{\xi_{i,k,w}^{\rm obs}}{\varepsilon_{\rm obs}}\Bigr)^4,
        \end{aligned}
    \end{equation}
    subject to the same constraints as in problem~\eqref{eq:mpc-problem-predprey}.
\end{itemize}
Both problems~\eqref{eq:mpc-problem-predprey} and~\eqref{eq:shielding-mpc-problem-predprey} are solved with prediction horizon \(N=5\) and the cost weights
\[
Q = \begin{bmatrix}
    100 & 0 \\ 0 & 100
\end{bmatrix}, \ R = \begin{bmatrix}
    0.1 & 0 \\ 0 & 0.1
\end{bmatrix}, \ R_0 = R_S \begin{bmatrix}
    100 & 0 \\ 0 &100
\end{bmatrix}.
\]
Safety distances are set to \(d_{\mathrm{safe}}=3.5\,\mathrm{m}\) (agent--agent) and \(r_{\mathrm{obs}}=2.5\,\mathrm{m}\) (agent--obstacle). The world radius is fixed to \(r_{\mathrm{world}}=20\,\mathrm{m}\).
Component-wise input and velocity bounds are
\[
\begin{aligned}
  &\text{prey:}\quad ||u_{i,k}||\le 1.0\,\mathrm{m/s^2},\;\; ||v_{i,k}||\le 2.5\,\mathrm{m/s},\\
  &\text{predator:}\quad ||u_{i,k}||\le 0.75\,\mathrm{m/s^2},\;\; ||v_{i,k}||\le 1.5\,\mathrm{m/s},
\end{aligned}
\]
for all three RL schemes. For the ROI-guided learning the radius of the ROI is fixed to \(r_{\mathrm{ROI}} = 10\,\mathrm{m}\).

Each RL scheme is trained in three different environmental layouts: 
\begin{itemize}
  \item \textbf{Layout 1:} Two predators, three preys, no obstacles.
  \item \textbf{Layout 2:} Same as \textbf{1}, plus three circular obstacles.
  \item \textbf{\mbox{Layout 3:}} Same as \textbf{2}, with episode termination upon any predator-predator or predator-obstacle collision. A maximal negative reward is assigned in the event of such collisions.
\end{itemize}
These scenarios range from simple pursuit to safety‐critical, obstacle‐dense tasks. At the start of each episode, predators, preys, and obstacles are spawned in random positions (subject to a minimum separation). Each episode runs for at most \(T=400\) steps and terminates early once all preys have been caught. A prey is considered caught once a predator comes within \(2\,\mathrm{m}\) of it. Preys follow a deterministic evasion heuristic: Each of them evaluates a polar grid of candidate positions and selects the point that maximizes its minimal distance to all predators. If the path toward that target is unsafe, the prey switches to reactive fleeing from the nearest predator, accelerating in the escape direction.

In all three MARL schemes, each predator’s observation vector concatenates its own normalized polar position and velocity; its own and its partner’s last action or target (also in normalized polar form); the other predator’s relative polar position and velocity; and, for each prey, a binary alive flag plus its relative polar position and normalized speed. When obstacles are enabled, their relative polar positions are simply appended to the same vector. This unified, fixed-length encoding ensures that the policy receives identical structured inputs across all schemes.

The reward function is defined as a constant step penalty:  
\begin{equation}
  r_t = -0.1, \qquad 
  \mathcal{R}^{\mathrm{ep}} = \sum_{t=0}^{T-1} r_t = -0.1\,T ,
  \label{eq:reward}
\end{equation}
where $\mathcal{R}^{\mathrm{ep}}$ denotes the cumulative reward over one episode for a single agent. 
This minimal design encourages faster captures while keeping the reward structure simple and comparable across schemes.

Training and subsequent execution are performed on a workstation running Ubuntu~24.04~LTS with an AMD~Ryzen\texttrademark~9~9950X (32 threads), 64\,GB RAM, and an NVIDIA~GeForce~RTX\texttrademark~5080 GPU. We use MARLlib’s MAPPO implementation~\cite{MARLlib} with shared rewards, a learning rate of \(5\times10^{-4}\), discount factor \(\gamma = 0.99\), batch size of 80\,000, and a time step of \(\Delta t = 1\,\mathrm{s}\). The policy/value networks are MLPs with four hidden layers of sizes \(512\!-\!512\!-\!256\!-\!256\). Each variant is trained for 20 million environment steps. The decentralized MPC problems are formulated in \texttt{CasADi}~\cite{Andersson2019} and solved online using the IPOPT interior-point solver \cite{IPOPT}.

\begin{figure*}[ht]      
  \centering
  \includegraphics[trim=0.35cm 0.35cm 0.28cm 0.12cm, clip, width=\linewidth]{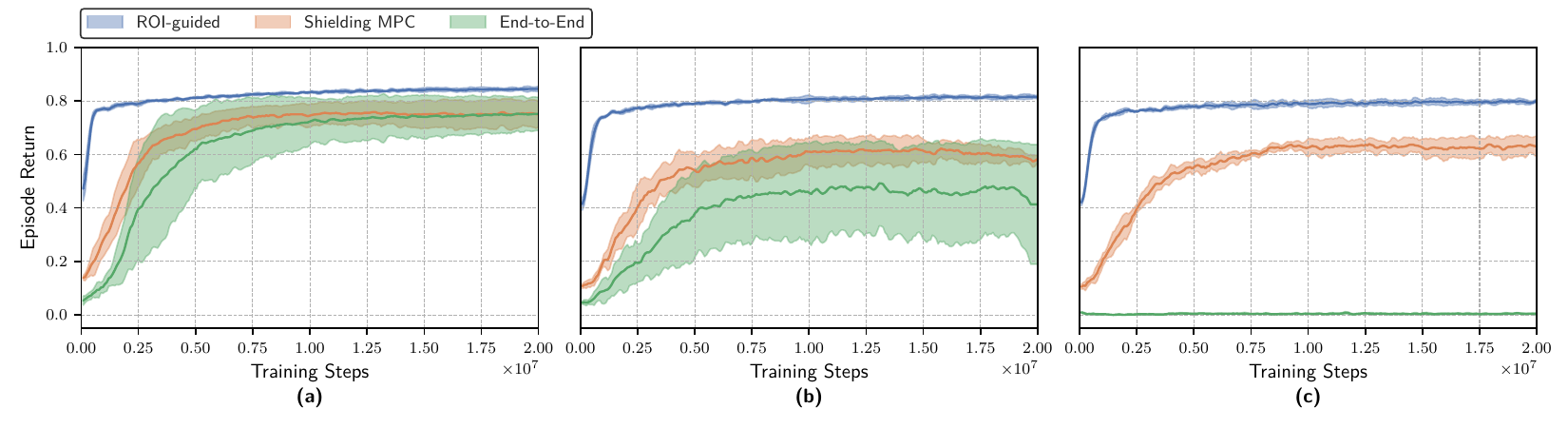}
  \caption{Training reward curves for the three MARL schemes, ROI-guided learning, End-to-End approach, and Shielding MPC, across the three evaluation layouts: \textbf{(a)} Layout 1: no obstacles, \textbf{(b)} Layout 2: obstacles only, and \textbf{(c)} Layout 3: with obstacles and episode termination upon any predator-predator or predator-obstacle collision. Each subplot shows the evolution of the episode reward over 20 million training steps.}
  \label{fig:Traininig Results}
\end{figure*}

The trainings results are given in Fig. \ref{fig:Traininig Results}. 
Across all three evaluation layouts, our ROI-guided learning approach with MPC not only converges the fastest and achieves the highest asymptotic reward (around 0.8 of the max. reward), but also begins training at a substantially higher reward level than both baselines. In contrast, both the End-to-End RL and the Shielding MPC baselines start near zero reward. The End-to-End method must learn both locomotion and tactics from scratch, while the Shielding MPC training corrects only locally unsafe actions without providing any structural trajectory guidance, leading to slower and less sample-efficient learning in early stages.

However, it is worth noting that both baselines exhibit similar learning curves under the absence of obstacles and eventually converge to moderately effective capture strategies. This could be expected, as the Shielding MPC only intervenes upon constraint violations and otherwise leaves the learned policy unchanged. This means that both End-to-End and Shielding MPC operate under nearly identical training conditions.

The difference in sample efficiency is particularly apparent when comparing the training steps required to reach \(80\,\%\) of the final convergence reward in averaged over all three environment layouts and summarized in Tab. \ref{tab:80percent}.
\begin{table}[ht]
  \centering
  \caption{Trainings steps required to reach \(80\%\) of the final convergence reward}
  \label{tab:80percent}
  \footnotesize
  \setlength{\tabcolsep}{4pt}
  \begin{tabularx}{\columnwidth}{l c c c}
    \toprule
     \textbf{Metric} & \textbf{\makecell[c]{ROI-\\ guided}} & \textbf{End-to-End} & \textbf{\makecell[c]{Shielding\\ MPC}} \\
    \midrule 
     Trainings steps required [millions]      & 2.288 & 7.125 & 4.039 \\
    \bottomrule
  \end{tabularx}
\end{table}

Notably, in the most challenging scenario (Layout 3), learning fails entirely for the End-to-End approach. Furthermore, the final reward levels achieved by both baselines remain significantly below those of the ROI-guided learning approach. Since the reward function (\ref{eq:reward}) penalizes capture time, a higher reward directly corresponds to faster and more efficient coordination. Additional performance metrics are reported in Tab.~\ref{tab:core_metrics_env}.

\begin{table}[ht]
  \centering
  \caption{Core performance metrics per environment and method (aggregated over 1{,}000 evaluation episodes)}
  \label{tab:core_metrics_env}
  \footnotesize
  \setlength{\tabcolsep}{4pt}
  \begin{tabularx}{\columnwidth}{l l c c c}
    \toprule
    \textbf{\makecell[l]{Env.\\ Layout}} & \textbf{Metric} & \textbf{\makecell[c]{ROI-\\ guided}} & \textbf{End-to-End} & \textbf{\makecell[c]{Shielding\\ MPC}} \\
    \midrule
    \multirow{2}{*}{1)} 
      & Full Capture Rate [\%]        & 100 & 98.89 & 98.76 \\
      & Avg.\ Capture Time [Steps]   & 65.02 & 98.45 & 96.21 \\
    \addlinespace
    \multirow{2}{*}{2)} 
      & Full Capture Rate [\%]        & 99.98 & 81.41 & 93.13 \\
      & Avg.\ Capture Time [Steps]   & 75.18 & 194.7 & 171.8  \\
    \addlinespace
    \multirow{3}{*}{3)} 
      & Full Capture Rate [\%]        & 99.4 & 0.8    & 96.57 \\
      & Avg.\ Capture Time [Steps]   & 75.25 & -    & 144.7 \\
      & \makecell[l]{Collisions per \\ 1{,}000 Episodes}                       &  6 &  986 &  6 \\
    \bottomrule
  \end{tabularx}
\end{table}

In Layout~1 and 2, all methods achieve high capture rates, but the ROI-guided policy consistently reaches the goal significantly faster. In Layout~3, End-to-End learning fails to generalize effectively, while the ROI-guided method maintains high capture rates and minimizes collisions. Shielding-MPC improves safety but remains less efficient in terms of capture time. 

We also evaluate how well the learned ROI-guided policy generalizes to randomized radii of the region of interest in Layout~3.  The policy trains with a fixed ROI radius \(r_{\mathrm{ROI}}=10\,\mathrm{m}\), while evaluation uses an ROI with randomized radii between \(5\,\mathrm{m}\) to \(15\,\mathrm{m}\). Across 1{,}000 episodes, the mean capture time is \(78.61\) steps, only marginally above the fixed-radius setting. Complete capture occurs in \(99.5\%\) of episodes, and only five collisions (predator-predator or predator-obstacle) have occurred across all runs.

\section{Discussion}
In our benchmark setting, continuous-control predator–prey with increasing safety demands, the ROI-guided approach consistently outperforms both End-to-End learning and the Shielding MPC approach in learning speed, final reward, and robustness under constraints.

\subsubsection*{Advantages of embedded MPC layer}
The internal MPC layer provides key benefits: Constraints are handled explicitly through optimization rather than implicitly through reward shaping, and the agent’s policy focuses on strategic intent rather than low-level control. By offloading trajectory tracking to MPC, the policy operates in a simplified decision space, improving sample efficiency and stability.

\subsubsection*{ROI guidance}
Introducing an MPC layer naturally increases algorithmic complexity, as the policy must now interface with a finite-horizon optimizer. However, by restricting the policy output to a structured region of interest, we simplify the learning task: The ROI reduces unnecessary exploration and narrows the search to semantically meaningful targets. Although the ROI introduces an additional design element, it primarily acts as a mechanism for incorporating prior knowledge. Specific ROI designs (e.g., Voronoi-based or adjacency-restricted) are context-dependent and can be adapted to different domains.

\subsubsection*{End-to-End learning in contrast}
While pure End-to-End learning may achieve similar performance given sufficient training and careful reward shaping, our results suggest that embedding a small, well-structured model-based core dramatically accelerates learning and enhances safety. Rather than replacing learning, this architecture guides it more efficiently through a meaningful structure.

\section{Conclusion}
The fusion of MPC and RL into a ROI-guided learning framework enables robust, sample-efficient multi-agent control under constraints. By structuring decision-making around interpretable tactical choices and relying on certified low-level execution, this approach contributes to safe learning in real-world systems. A promising path for future research is to exploit its modularity by applying the framework across a wider range of domains, such as other multi-agent or even single-agent systems.

\bibliography{refs}
\end{document}